# Rare $b$ Decays

*Stephen Playfer and Sheldon Stone*
Physics Department
Syracuse University
Syracuse, New York 13244-1130

**Abstract**

Rare $b$ decays provide a unique opportunity to measure Standard Model parameters and probe beyond the Standard Model. We review here the experimental progress made in measuring these decays, and the importance of future measurements, including the possible observation of CP violation.

## I. Introduction

The dominant decays of the $b$ quark are charged current couplings via a $W^-$ to a $c$ quark as shown in Figure 1(a). There are also rare decays to a $u$ quark. Observation of these decays has led to measurements of the Cabibbo-Kobayashi-Maskawa (CKM) matrix elements $|V_{cb}|$ and $|V_{ub}|$ [1].

The $b$ quark can make transitions in other ways. $B^0 - \bar{B}^0$ mixing, the process where a particle changes into its antiparticle, occurs via a "box" diagram with virtual $W$ bosons and $t$ quarks inside the box [Figure 1(b)]. The box diagram gives rise to large fractions of mixed events: 17% for $B^0$ and 50% for $B_s$ mesons.

Flavor-changing neutral currents lead to the transitions $b \to s$ and $b \to d$. These can be described in the Standard Model by one-loop diagrams, known as "penguin" diagrams, where a $W^-$ is emitted and reabsorbed [2]. The first such process to be observed was $b \to s\gamma$, described by the diagram in Figure 1(c), where the $\gamma$ can be radiated from any charged particle line. Another process which is important in rare $b$ decays is $b \to sg$, where $g$ designates a gluon radiated from a quark line [Figure 1(d)]. A third example of such processes is the transition $b \to s\ell^+\ell^-$ which can occur through the diagrams shown in Figures 1(e) and 1(f). We consider the loop processes shown in Figures 1(c)-(f) to be among the most interesting and important rare $b$ decays.

The decay amplitudes for the diagrams shown in Figure 1 are proportional to the CKM matrix elements present at each vertex. For the loop diagrams there are additional factors of $\alpha$ if a $\gamma$ is radiated and $\alpha_s$ if a gluon is radiated, as well as a kinematic factor which is a function of $(m_q/m_W)^2$. Since the heaviest quark is the top quark, it is usually the amplitude involving the top quark that dominates in decays via loop diagrams.

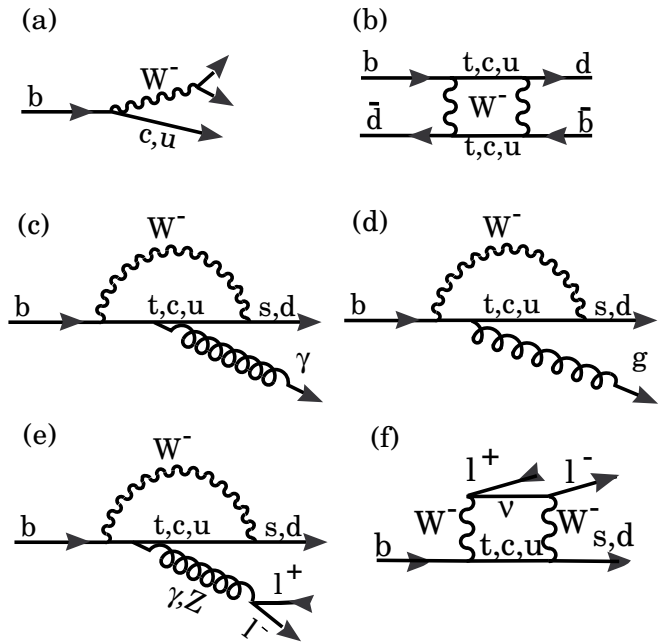

Figure 1: Feynman Diagrams for $b$ Decays

$B$ meson decays presently provide the only experimental evidence for penguin decays. Although neutral kaons have long been known to mix, penguin contributions to kaon decay are hard to identify since the $s \to d$ transition leads to the same final states as the $s \to u$ transition [3]. Only the dilepton final states can give direct evidence for penguins in kaon decay. The channels $K^+ \to \pi^+ \nu \bar{\nu}$ and $K^0 \to \pi^0 e^+ e^-$ are the most promising, but the predicted branching ratios are small [4], and thus far they have not been observed.

Loop diagrams in charm decays are suppressed either by CKM matrix elements or by small values of $(m_q/m_W)^2$, since the heaviest available quark is the $b$ quark. The mixing diagram for $D^0$ decay is proportional to $|V_{cb}V_{ub}|^2 \times (m_b/m_W)^2$ for $b$-quark exchange and to $|V_{cs}V_{us}|^2 \times (m_s/m_W)^2$ for $s$-quark exchange. The same kind of suppression factors apply in the case of the penguin diagrams. As a result decays such as $D \to \rho \gamma$ are expected to be dominated by long distance contributions such as rescattering from the related hadronic decay $D \to \rho\rho$. In the case of mixing a long distance effect would be $D^0 \to K^+K^- \to \bar{D}^0$. If these long distance effects are not too large, rare charm decays may be sensitive to non-Standard Model effects, since the Standard Model predictions for the loop diagrams are so small [5].

It is expected that CP violation will be significant in rare $b$ decays. Charge conjugation, C, changes a particle to an anti-particle, while parity, P, changes left-handed particles to right-handed or vice-versa. In 1964 it was found that the combined operation, CP, showed an asymmetry in neutral kaon decays [6]. CP violation is a necessary ingredient in explaining why our local position in the Universe consists of matter rather than anti-matter, and thus why we exist. It is of great importance to find out whether or not the Standard Model can quantitatively describe CP violation in the $B$ system.

In the parameterization of Wolfenstein [7], the CKM matrix can be described by four independent parameters $\lambda$, $A$, $\rho$ and $\eta$. The matrix is given in equation (1).

$$V_{ij} = \begin{pmatrix} 1 - \lambda^2/2 & \lambda & A\lambda^3(\rho - i\eta) \\ -\lambda & 1 - \lambda^2/2 & A\lambda^2 \\ A\lambda^3(1 - \rho - i\eta) & -A\lambda^2 & 1 \end{pmatrix} \quad (1)$$

The $\lambda$ and $A$ parameters have been measured in semileptonic decays of $s$ and $b$ quarks. Although $\eta$ and $\rho$ have not been determined separately, constraints on these parameters are given by measurements of the $\epsilon$ parameter describing CP violation in $K_L^0$ decay, and by $B^0 - \bar{B}^0$ mixing and semileptonic $b \to X\ell\nu$ decays. An analysis of the allowed parameter space is shown in Figure 2 [8]. Overlaid on the figure is a triangle that results from the requirement $V_{3k}V_{1k}^\dagger = 0$, i.e. that the CKM matrix be unitary. Measurements of CP violation in $B$ decays can, in principle, determine each of the angles $\alpha$, $\beta$ and $\gamma$ of this triangle independently.

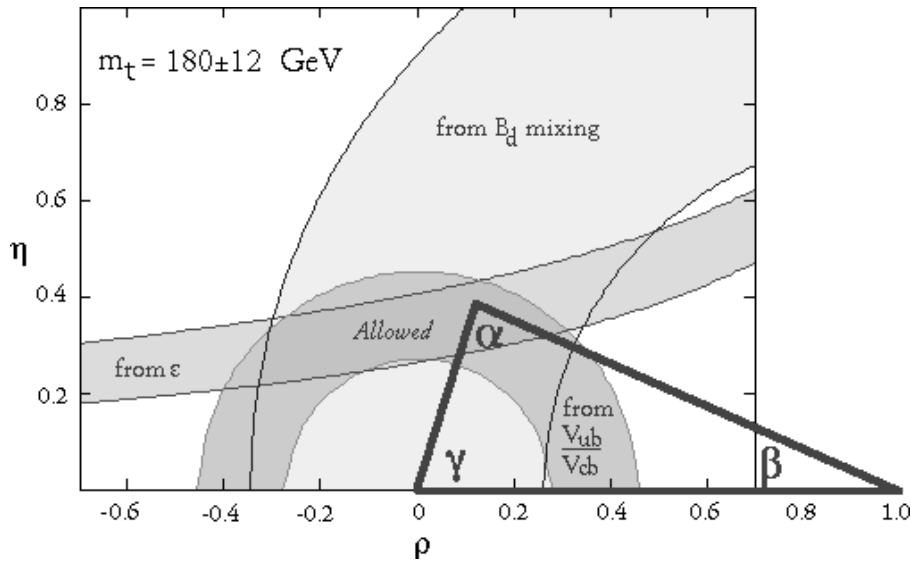

Figure 2: The CKM triangle overlaid upon constraints in the $\rho - \eta$ plane, from measured values of $V_{ub}/V_{cb}$, $B^0 - \bar{B}^0$ mixing and $\epsilon$ in the $K^0$ system. The allowed region is given by the intersection of the three bands.

In this paper we will review the experimental data on rare $b$ decays and compare it with the Standard Model predictions. Following this we will discuss the sensitivity of the data to extensions of the Standard Model. Finally we discuss the importance of CP violation and the propects for experimental measurements.

# II. $B^0 - \bar{B}^0$ Mixing

The transformation of a $B^0$ meson into a $\bar{B}^0$ meson can occur via the diagram shown in Figure 1(b). As this is not the main topic of this paper we give only a brief summary here and refer the reader to an excellent review for more details [9]. The variable that is measured by experiments is $x \equiv \Delta M/\Gamma$, where $\Delta M$ is the mass difference between the light and heavy neutral $B$ mesons. The CKM elements are related to $x$ via

$$x = \frac{G_F^2}{6\pi^2} B_B f_B^2 m_b \tau_B |V_{tb}^* V_{td}|^2 F\left(\frac{m_t^2}{M_W^2}\right) \eta_{QCD}, \qquad (2)$$

where $G_F$ is the Fermi constant. The constant $B_B$ and the $B$ meson decay constant $f_B$ have been calculated theoretically, but the large uncertainties in these calculations limit the ability to extract the CKM element $|V_{td}|$ from the measurement of $x$. To determine $x$ experiments either measure the ratio of mixed events to total events integrated over time (ARGUS and CLEO), or they measure the explicit time dependence (ALEPH and OPAL). The extracted $x$ values are shown in Table 1.

Table 1: $x = \Delta M/\Gamma$ Values from $B_d^0$ mixing measurements

| Experiment | $x$ |
|---|---|
| CLEO[10] | 0.65±0.10 |
| ALEPH[11] | 0.76±0.12 |
| OPAL[12] | 0.73±0.14 |
| ARGUS[13] | 0.75±0.15 |
| AVERAGE | 0.71±0.06 |

The band in Figure 2 is derived from equation (2) by assuming $B_B = 1$, and taking an $f_B$ range of 160-240 MeV that corresponds to recent theoretical estimates.

The fraction of mixed events is given by

$$\chi = \frac{x^2}{2(1+x^2)} \qquad (3)$$

The measurements of $x$ correspond to a $\chi$ value of 17% for $B^0$ events. There are also predictions and experimental limits on mixing in the $B_s$ system indicating that the mixing has an almost maximal value of 50% [9].

# III. Observation of Radiative Penguin Decays

Figure 1(c) depicts the process $b \to s\gamma$, where the photon can be radiated by any charged object in the diagram. This decay is uniquely described in the Standard

Model by a "penguin" diagram, with corrections from other diagrams, often called "long distance" effects, estimated to be only a few percent (see Section V(A) for a detailed discussion). The inclusive process $b \to s\gamma$ leads to many exclusive final states where the $s$ quark hadronizes with the spectator quark. Angular momentum conservation forbids the decay $B \to K\gamma$, but it is expected that $K^*(892)\gamma$ will be a significant fraction of the inclusive rate. The remaining inclusive rate comes from higher mass $K^*$ resonances and non-resonant $K(n\pi)$ final states. There are large variations among the theoretical predictions for the fraction of $b \to s\gamma$ that hadronizes as $B \to K^*\gamma$.

## A) Observation of $B \to K^*\gamma$

The first successful search for $b \to s\gamma$ by the CLEO collaboration was for the exclusive $K^*\gamma$ final state [14]. This is much easier than trying to measure the inclusive branching ratio for $b \to s\gamma$, because the final state is completely kinematically constrained, and the analysis is similar to that used for reconstructing hadronic $B$ meson final states at the $\Upsilon(4S)$ [15]. Neutral clusters in a CsI calorimeter are selected with energies between 2.1 and 2.9 GeV, if they have a shower shape consistent with a single $\gamma$, and if they cannot be combined with another $\gamma$ to form a $\pi^0$. The $K^*(892)$ candidates are searched for in three channels: $K^{*0} \to K^+\pi^-$, $K^{*-} \to K^-\pi^0$ and $K^{*-} \to K^0\pi^-$. If the energy sum of the $K^*$ and the $\gamma$ is within 75 MeV of the known beam energy $E_{beam}$, then the beam constrained invariant mass

$$m_B = \sqrt{E_{beam}^2 - \left(\overrightarrow{P_{K^*}} + \overrightarrow{P_\gamma}\right)^2} \qquad (4)$$

is plotted for each candidate event and an excess is looked for at the known $B$ meson mass.

The difference in shape between jetlike continuum events and spherical $B\bar{B}$ events is exploited by making cuts on several event shape variables to suppress the continuum background. The most useful variables are the angle of the thrust axis of the rest of the event relative to the candidate thrust axis ($\cos\theta_T$), the second Fox-Wolfram moment ($R_2$) [16], and the sum of the momenta in a 90° cone perpendicular to the candidate axis ($s_\perp$) [14, 17]. There is a significant background due to initial state radiation (ISR). To suppress this background the events are transformed to the rest system of the $e^+e^-$ following the emission of the photon. In this primed frame the variables $\cos\theta'_T$ and $R'_2$ are recalculated.

In $1.4 fb^{-1}$ of $\Upsilon(4S)$ data there are eight $K^{*0}\gamma$ and five $K^{*-}\gamma$ candidates within 6 MeV of $M_B$. The continuum background level is one event in each of $K^{*0} \to K^+\pi^-$ and $K^{*-} \to K^-\pi^0$, and zero in $K^{*-} \to K^0\pi^-$ where there are two candidates. This is a clear signal for the decay $B \to K^*\gamma$ (Figure 3). The yields of $B^0 \to K^{*0}\gamma$ and $B^- \to K^{*-}\gamma$ are consistent. If the relative fractions of $B^-$ and $B^0$ produced at the $\Upsilon(4S)$ are assumed to be equal, the average branching ratio is $(4.5 \pm 1.5 \pm 0.9) \times 10^{-5}$.

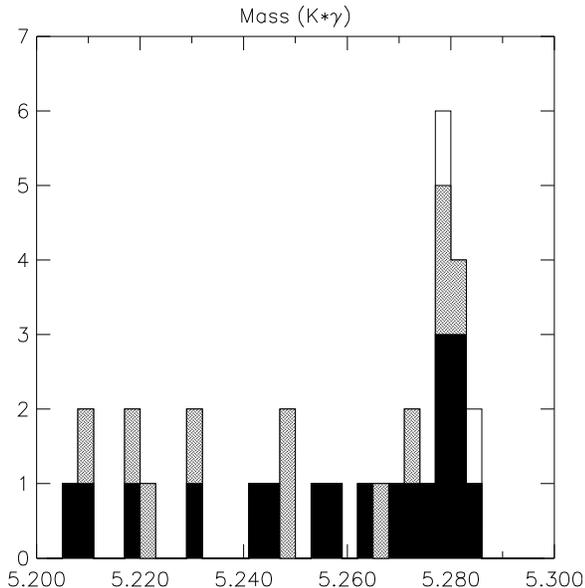

Figure 3: Beam constrained mass distribution in GeV for the $B \to K^*\gamma$ candidates, dark shaded $K^{*0} \to K^+\pi^-$, light shaded $K^{*-} \to K^-\pi^0$, unshaded $K^{*-} \to K^0\pi^-$

## B) Measurement of Inclusive $b \to s\gamma$

Recently, the CLEO collaboration has also made the first measurement of the inclusive $b \to s\gamma$ branching ratio [18]. The signature for the inclusive process is a photon with energy between 2.2 and 2.7 GeV. This region contains 75-90% of the signal according to calculations that include the smearing due to the Fermi motion of the quarks in the $B$ meson, and the motion of a $B$ meson produced at the $\Upsilon(4S)$.

There are large backgrounds to the inclusive signal from continuum jets ($e^+e^- \to q\bar{q}$) and initial state radiation (ISR). These backgrounds are suppressed by two methods: a shape variable analysis using a neural network, and a $B$ reconstruction analysis. After these cuts have been made, the remaining continuum background is subtracted using scaled off-resonance data. There are also small backgrounds from other $B$ decays, mostly consisting of photons from $\pi^0$ and $\eta$ decays that survive a $\pi^0(\eta)$ mass cut because the other photon was not found. As a first approximation these are taken from a Monte Carlo simulation, but then a correction is made for any differences that are observed between the $\pi^0$ and $\eta$ spectra measured in data, and those predicted by the Monte Carlo. This takes into account any omissions in the Monte Carlo (e.g. $b \to sg$).

The neural network analysis uses a set of eight variables defining the event shape. The variables $R_2$, $s_\perp$, $R_2'$ and $\cos\theta_T'$ are as defined in the previous section. In addition the energies in $20°$ and $30°$ cones parallel and antiparallel to the $\gamma$ direction are used. The energies in the "away" cones relative to the $\gamma$ are found to be particularly useful in discriminating against both $q\bar{q}$ and ISR backgrounds. However since the eight variables are highly correlated, and none of them has clear discriminating power compared to the others, they are combined into a joint variable, $r$, which tends to $+1$ for signal, and $-1$ for continuum background. A neural net is used for this purpose since it is the best method of taking into account the correlations between the shape variables.

The $B$ reconstruction analysis combines the high energy photon with a candidate $X_s$ system, where $X_s$ contains either a $K_s \to \pi^+\pi^-$ or a charged track consistent with a kaon, and an additional 1-4 pions, of which one may be a $\pi^0$. To be accepted the reconstructed decay candidate must satisfy a thrust axis cut, $|\cos\theta_T| < 0.7$, and a $\chi^2$ cut on the combined $\Delta E$ and $m_B$ information. If there is more than one candidate per event, the one with the smallest $\chi^2$ is selected. The reconstruction ambiguities usually have the same high energy photon, but different $X_s$ systems. This is not important if the method is used only to suppress continuum background, and no attempt has been made to obtain the corrected $X_s$ mass distribution in the CLEO analysis. Figure 4 shows the apparent $X_s$ mass distribution, with a fit indicating the presence of a large component from $K^*(892)$. With larger data samples it will be possible to study the $X_s$ mass distribution and obtain additional information about the exclusive decay modes that contribute.

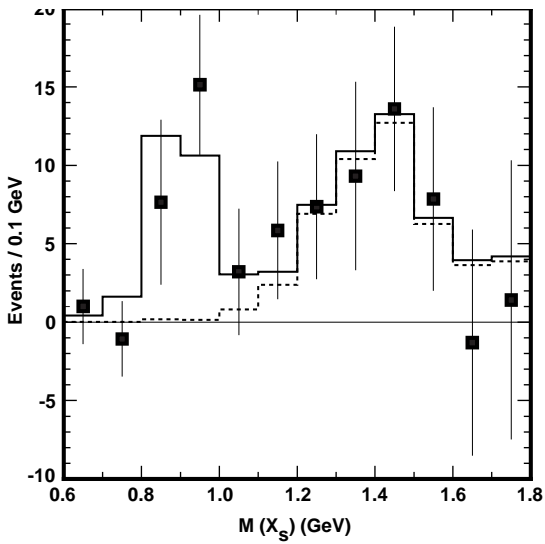

Figure 4: Apparent $X_S$ mass distribution from the $B$ reconstruction analysis. The solid curve is a fit to the expected distribution from a spectator model. The dashed curve shows the non-K*(892) component of the fit.

The two methods for suppressing continuum are complementary. The neural net has high efficiency (32%) but modest background suppression, whereas the $B$ reconstruction method has low efficiency (9%), but suppresses the background by an additional factor of 14. According to Monte Carlo studies they should be equally sensitive and only slightly correlated with each other. Figure 5 shows the photon en-ergy spectra from the two analyses. In Figures 5(b) and (d) the signal shape is taken from a spectator model prediction [19]. There is a small excess above the kinematical endpoint in Figure 5(b) that is attributed to a statistical fluctuation in the continuum background. The measured branching ratios are $\mathcal{B}(b \to s\gamma) = (1.88 \pm 0.74) \times 10^{-4}$ from the event-shape analysis and $\mathcal{B}(b \to s\gamma) = (2.75 \pm 0.67) \times 10^{-4}$ from the $B$ reconstruction analysis. The average result, after taking into account the small correlations between the two analyses, is $\mathcal{B}(b \to s\gamma) = (2.32 \pm 0.57 \pm 0.35) \times 10^{-4}$, where the first error is statistical, and the second systematic. Details of the contributions to the systematic error can be found in [20].

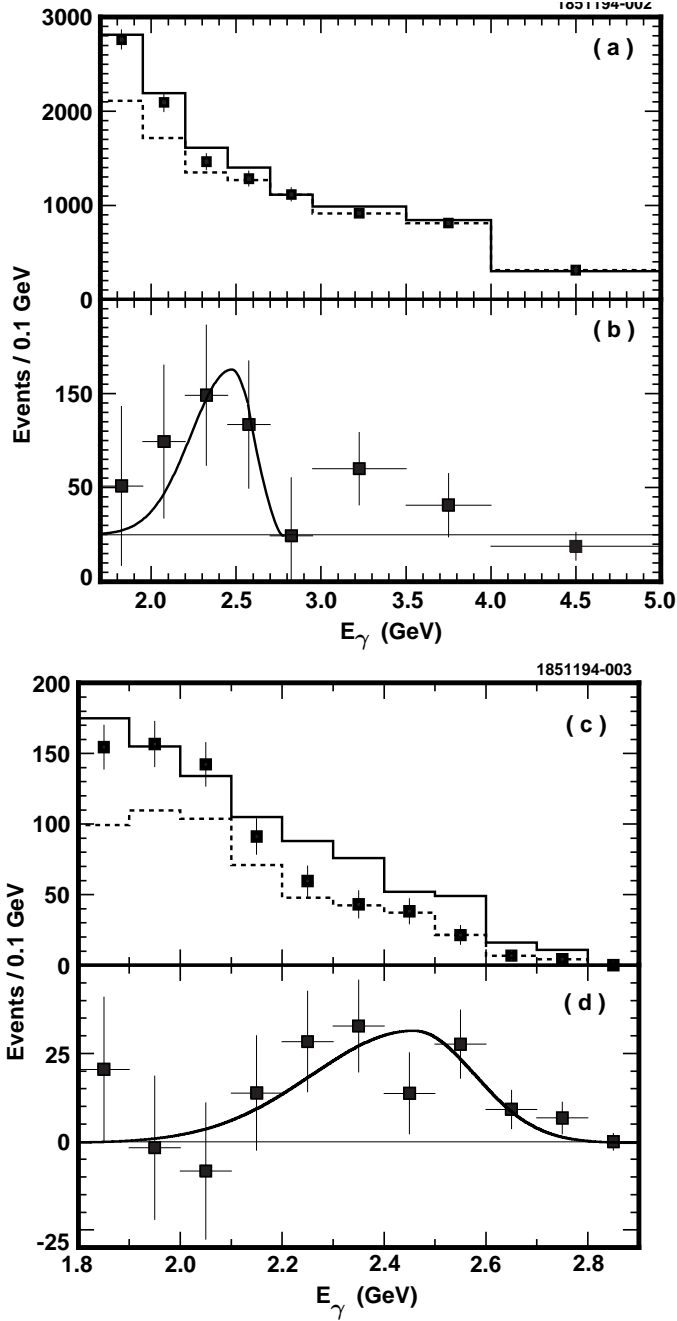

Figure 5: Photon energy spectra from the neural net analysis, (a) & (b), and from the $B$ reconstruction analysis, (c) & (d). In (a) & (c) the on resonance data are the solid lines, the scaled off resonance data are the dashed lines, and the sum of backgrounds from off resonance data and $b \to c$ Monte Carlo are shown as the square points with error bars. In (b) & (d) the backgrounds have been subtracted to show the net signal for $b \to s\gamma$. The solid lines are fits of the signal shape using a spectator model prediction.

## IV.  Theory of Radiative Penguin Decays

### A)  Standard Model Prediction for $b \to s\gamma$

The partial decay width for $b \to s\gamma$ is given by [21, 22]:

$$\Gamma(b \to s\gamma) = \frac{\alpha G_F^2 m_b^5}{128\pi^4} |V_{ts}^* V_{tb} C_7^{eff}(\mu)|^2 \tag{5}$$

Since the quark mass, $m_b$, is not well known, the $m_b^5$ dependence is removed by normalizing to the decay rate for $b \to c\ell\nu$:

$$\frac{\Gamma(b \to s\gamma)}{\Gamma(b \to c\ell\nu)} = \frac{|V_{ts}^* V_{tb}|^2}{|V_{cb}|^2} \frac{\alpha}{6\pi g(m_c/m_b)} |C_7^{eff}(\mu)|^2 \tag{6}$$

where the factor $g(m_c/m_b)$ corrects for phase space. In these expressions $C_7^{eff}(\mu)$ is an effective coefficient of the electromagnetic loop operator:

$$\mathcal{O}_7 = \frac{e}{8\pi^2} m_b \bar{s}_\alpha \sigma^{\mu\nu} (1 + \gamma_5) b^\alpha F_{\mu\nu} \tag{7}$$

The value of $C_7$ can be calculated perturbatively at the mass scale $\mu = M_W$. The explicit expression for $C_7(M_W)$ as a function of $(m_t^2/M_W^2)$ can be found in [23]. The evolution from $M_W$ down to a mass scale $\mu = m_b$ introduces large QCD corrections. These are calculated using an operator product expansion based on an effective Hamiltonian:

$$\mathcal{H}_{eff}(b \to s\gamma) = -2\sqrt{2} G_F V_{ts}^* V_{tb} \sum_{i=1}^{8} C_i(\mu) \mathcal{O}_i(\mu) \tag{8}$$

Renormalization of the coefficients, $C_i$, and operator mixing, lead to a value of $C_7^{eff}(\mu)$ significantly larger than $C_7(M_W)$ [23]. This increases the predicted rate for $b \to s\gamma$ by a factor of 2-3.

Evidently the prediction for the rate is very sensitive to the QCD corrections. The leading log calculation is uncertain to about 25%, primarily because it is unclear at which renormalization scale, $\mu$, the effective coefficient, $C_7^{eff}(\mu)$, resulting from the operator product expansion, should be evaluated. Values between $\mu = \frac{1}{2}m_b$ and $\mu = 2m_b$ have been suggested. A next-to-leading order calculation requires the evaluation of additional two-loop diagrams, as well as some three-loop diagrams. It is hoped that these calculations can be done, since they are expected to reduce the uncertainty in the Standard Model prediction to about 10%.

## B)   Comparison between $b \to s\gamma$ Experiment and Theory

The leading log prediction for $\mathcal{B}(b \to s\gamma)$ is $(2.8 \pm 0.8) \times 10^{-4}$ [22, 23]. If the next-to-leading order terms that have been calculated are included they tend to reduce the prediction to about $1.9 \times 10^{-4}$ [24]. Both these predictions are in excellent agreement with the experimental result of $(2.3 \pm 0.6 \pm 0.4) \times 10^{-4}$. Since the theoretical uncertainties are dominated by the choice of the renormalization scale, $\mu$, it is difficult to obtain useful constraints on other Standard Model parameters such as $m_t$ and $V_{ts}$. The combined CDF and D0 measurement of $m_t = (180 \pm 12)$ GeV [25] is well within the range required for consistency with $b \to s\gamma$. Ali et al. [26] have set bounds on $V_{ts}$:

$$0.62 < \frac{|V_{ts}|}{|V_{cb}|} < 1.10 \qquad (9)$$

This ratio is expected to be one if the CKM matrix is unitary.

Table 2: Predictions for the ratio of $B \to K^*\gamma$ to $b \to s\gamma$.

| Author(s) | Reference | Method | $B \to K^*\gamma$ Fraction |
|---|---|---|---|
| Altomari | [27] | Spectator Quark Model | 4.5% |
| Deshpande & Trampetic | [28] | Relativistic Quark Model | 6 - 14% |
| Aliev *et al* | [29] | QCD Sum Rules | 39% |
| Ali & Greub | [19] | Spectator Quark Model | $(13\pm3)\%$ |
| O'Donnell & Tung | [30] | Heavy Quark Symmetry | 10% |
| Ball | [31] | QCD Sum Rules | $(20\pm6)\%$ |
| Atwood & Soni | [32] | Bound State Resonances | 1.6 - 2.5% |
| Bernard, Hsieh & Soni | [33] | Lattice QCD | $(6.0\pm1.2\pm3.4)\%$ |
| UKQCD collaboration | [34] | Lattice QCD | 15 - 35% |

The fraction of the inclusive $b \to s\gamma$ rate hadronizing as $B \to K^*\gamma$ depends on the $B \to K^*$ form factor. This has been calculated by many authors using either QCD sum rules, Lattice QCD, or Heavy Quark Effective Theory (HQET). Table 2 summarizes these predictions for the ratio of $B \to K^*\gamma$ to $b \to s\gamma$. It can be seen that the predictions range from a few percent to 40%. The data suggest a value of $(21 \pm 7)\%$ for this ratio, which is not accurate enough to limit the range of acceptable form factor models. It has been suggested by Isgur [35] that the discrepancies between the models could be resolved by using the measured $D \to K^*$ form factors as a basis for calculating all heavy to light quark form factors. Larger data samples will make it possible to distinguish between the predictions in Table 2 and improve our understanding of the $B \to K^*$ form factor. Until accurate predictions are available for the exclusive measurements, they are not as useful as the inclusive measurement for constraining the Standard Model or new physics.

## C) Effect of Extensions of the Standard Model on $b \to s\gamma$

The measurement of $b \to s\gamma$ has inspired a large number of theoretical investigations of extensions of the Standard Model that could lead to significant changes in the predicted rate for $b \to s\gamma$. These studies use the upper and lower limits (95% C.L.):

$$1.0 \times 10^{-4} < \mathcal{B}(b \to s\gamma) < 4.2 \times 10^{-4} \tag{10}$$

to constrain the allowed parameter space of the Standard Model extension being considered. Among the most widely discussed models are Higgs doublets, Supersymmetry, anomalous $WW\gamma$ couplings, and anomalous top quark couplings. We give a brief summary of these cases below. For investigations into other theoretical ideas such as leptoquarks, a fourth generation and left-right symmetric models the reader is referred to the review article by Hewett [36].

In two-Higgs doublet models there is a charged Higgs that can be inserted into the loop instead of the $W$ boson. There are two models for the couplings of the Higgs doublets to the quarks, depending on how the fermion masses are generated. In both cases the free parameters are the charged Higgs mass, $M_{H^+}$, and the ratio of the doublet vacuum expectation values, $\tan\beta$. With Model I couplings, the $b \to s\gamma$ rate is enhanced at low $\tan\beta$, suppressed for values of $\tan\beta$ between 0.5 and 1.0, and is rather insensitive to large $\tan\beta$. Model II couplings always enhance the $b \to s\gamma$ rate. In this case the experimental upper limit requires $M_{H^+}$ to be at least 240 GeV even for large values of $\tan\beta$ [37].

Supersymmetry introduces many additional particles that can appear inside the loop. In the limit of exact supersymmetry these additional contributions cancel the Standard Model contribution, and $b \to s\gamma$ does not occur at all. In supersymmetric models there are charged Higgs bosons with Model II type couplings that enhance the rate for $b \to s\gamma$. Contributions in which down type squarks and either neutralinos or gluinos are inserted into the loop are usually found to be negligible. However, there are significant contributions when an up type squark and a chargino are inserted in the loop. There are several recent analyses of the size and sign of the chargino contributions relative to the Standard Model and charged Higgs contributions. It appears that there are some regions of the parameter space where the supersymmetric model predicts a rate comparable to or below the Standard Model, even for small values of $M_{H^+}$. This requires a small stop quark mass, a large value of $\tan\beta$, and a higgsino mass parameter $\mu < 0$ [38].

## D) Constraints on Anomalous Couplings

The existence of anomalous couplings at the $WW\gamma$ vertex can be constrained by tree-level processes such as $e^+e^- \to W^+W^-$ and $p\bar{p} \to W\gamma$, and by loop diagrams in processes such as $b \to s\gamma$ [39]. The anomalous couplings are described by two parameters, $\lambda$ and $\Delta\kappa$, which are zero in the Standard Model, but can acquire non-zero values in some extensions of the Standard Model. They modify the value of $C_7(M_W)$, and hence the predicted rate for $b \to s\gamma$:

$$C_7(M_W) = C_7(M_W)^{SM} + A_1 \Delta\kappa + A_2 \lambda \qquad (11)$$

The coefficients $A_1$ and $A_2$ are functions of $(m_t^2/M_W^2)$. Since $A_1$ is larger than $A_2$ for $m_t = 180$ GeV, $b \to s\gamma$ is three times more sensitivity to $\Delta\kappa$ than to $\lambda$. Figure 6 shows the bounds that can be set on $\lambda$ and $\Delta\kappa$ from existing data. The limits from $b \to s\gamma$ are complementary to the limits obtained at the Tevatron [40]. Large positive and negative values of $\Delta\kappa$ and $\lambda$ are excluded by the upper limit on $b \to s\gamma$. In the region around $\Delta\kappa = -1$ there is a complicated interference between the three terms in equation (11). This leads to the exclusion of a narrow band by the lower limit on $b \to s\gamma$.

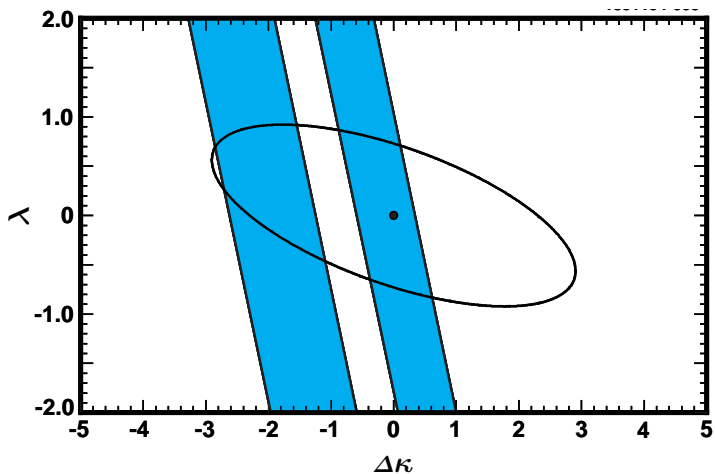

Figure 6: Limits on anomalous $WW\gamma$ couplings. The shaded regions are allowed by the $b \to s\gamma$ measurement. The region between the shaded regions is excluded by the lower limit, the outer unshaded regions by the upper limit. The ellipse shows the limit obtained at the Tevatron by the D0 experiment (CDF has a similar limit).

Anomalous top quark couplings have also been considered [41]. The first possibility is that there are anomalous $tt\gamma$ couplings in analogy to the $WW\gamma$ case considered above. Once again this would modify $C_7$ through two additional parameters. There is also the possibility of anomalous gluon couplings to the top quark that would modify $C_8$, but the constraints on these couplings from $b \to s\gamma$ are found to be rather weak.

Finally there is the interesting point that $b \to s\gamma$ probes the V-A structure of the $tbW$ and $tsW$ couplings [42].

## V. Searches for Other Radiative Penguin Decays

### A) The Decay $B \to \rho\gamma$

It was suggested by Ali [43] that the ratio of CKM elements $|V_{td}|/|V_{ts}|$ could be extracted from a measurement of:

$$\frac{\mathcal{B}(B^- \to \rho^-\gamma)}{\mathcal{B}(B^- \to K^{*-}\gamma)} = \frac{\mathcal{B}(B^0 \to \rho^0\gamma) + (B^0 \to \omega\gamma)}{\mathcal{B}(B^0 \to K^{*0}\gamma)} = \frac{|V_{td}|^2}{|V_{ts}|^2}\xi\Omega \qquad (12)$$

where $\Omega$ corrects for phase space, and $\xi$ corrects for SU(3) symmetry breaking.

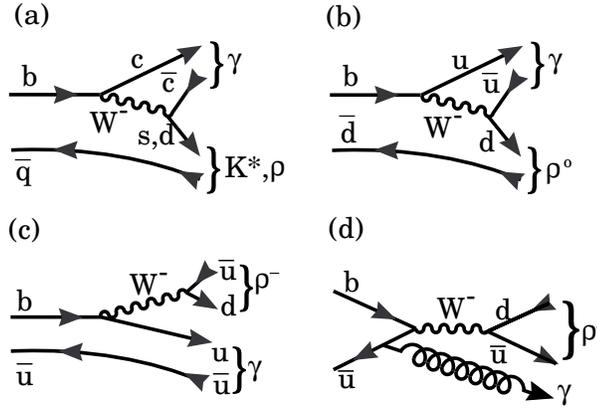

Figure 7: Non-penguin contributions to radiative decays. (a) Color suppressed diagram for $B \to K^*(\rho)\gamma$ with a $c\bar{c}$ intermediate state such as a $\psi$. (b) Color suppressed diagram for $B^0 \to \rho^0\gamma$ with a $u\bar{u}$ intermediate state such as a $\rho$. (c) Tree level diagram for $B^- \to \rho^-\gamma$ with a $u\bar{u}$ intermediate state such as a $\rho$. (d) Annihilation diagram for $B^- \to \rho^-\gamma$ where the $\gamma$ can be radiated from any of the lines.

Equation (12) is only valid if contributions other than the top-quark loop can be neglected in both decay modes. According to Soni [44] there are significant differences between the long-distance contributions to $b \to s\gamma$ and $b \to d\gamma$. Examples of such additional diagrams are shown in Figure 7. A recent estimate of the long distance contributions from virtual $\psi$ and $\rho$ mesons is $< 10\%$ for both $b \to s\gamma$ and $b \to d\gamma$ [45]. However, there is one contribution from an annihilation diagram (Figure 7(d)), that is predicted to be significant by Eilam et al [46], and may be as much as 60% of the top-quark loop. Note that this annihilation diagram includes the contribution from Figure 7(c) via rearrangement of the quark lines. Since this annihilation diagram only applies to $B^-$ decays it is expected that $\Gamma(B^- \to \rho^-\gamma)$ is different from $\Gamma(B^0 \to \rho^0\gamma)$.

Deshpande et al. [45] also discuss contributions from the $u$ and $c$ quark loops to $b \to d\gamma$. These contributions are larger than in $b \to s\gamma$ and may be as much as 20% of the top-quark loop. This would again complicate the extraction of $V_{td}$ from Equation (12).

CLEO has made a preliminary search for $B^- \to \rho^-\gamma$, $B^0 \to \rho^0\gamma$ and $B^0 \to \omega\gamma$ [47]. A data sample of $2.0 fb^{-1}$ at the $\Upsilon(4S)$ results in upper limits between 1.0 and $2.5\times 10^{-5}$ for the three modes. This corresponds to a limit on the ratio in equation (12) of 0.34 at 90% confidence level. The search is beginning to be background limited. In $\omega\gamma$ the background is primarily from the continuum, whereas in $\rho^-\gamma$ and particularily $\rho^0\gamma$ there is significant feeddown from misidentified $K^*\gamma$ events. Future detectors with better particle identification will be able to suppress this feeddown [48], but the continuum background may still be a problem.

## B) Searches for $b \to s\ell^+\ell^-$

The process $b \to s\ell^+\ell^-$ occurs through a loop diagram with a virtual $\gamma$ or Z boson (Figure 1(e)), or through a box diagram containing two W bosons (Figure 1(f)). In addition the hadronic decays $B \to \psi^{(\prime)} K^{(*)}$ contribute to the related exclusive decays $B \to K^{(*)}\ell^+\ell^-$ through the secondary decays $\psi^{(\prime)} \to \ell^+\ell^-$. A full understanding of $b \to s\ell^+\ell^-$ has to include both the short distance contributions from the loop and box diagrams, and the long distance contributions from the $\psi$ decays, and the interference between them [26, 49].

At low dilepton masses the dominant contribution from the virtual $\gamma$ can be directly related to $b \to s\gamma$. There are also sharp peaks from the $\psi$ contributions at $m_\psi$ and $m_{\psi'}$ which can be directly related to the measurements of the exclusive hadronic decays. At high dilepton masses the Z and box contributions are expected to be important, as are possible additional contributions from other heavy mass particles. The interference between the various diagrams can be studied by measuring the shape of the dilepton mass spectrum, and by measuring the lepton-pair asymmetry.

The high dilepton mass range has been studied at hadron colliders where there is a good signature for dimuon pairs. The first search for events with dimuon masses between 3.9 and 4.4 GeV was performed by the UA1 experiment [50]. They found upper limits of $5.0 \times 10^{-5}$ for the inclusive process $b \to s\mu^+\mu^-$, and $2.3 \times 10^{-5}$ for the exclusive channel $B^0 \to K^{*0}\mu^+\mu^-$. Both these limits should be interpreted as referring only to the short distance contributions from the loop and box diagrams, since there is an extrapolation to the remainder of the phase space under the assumption that the long distance contributions are negligible above $m_{\psi'}$. Recently the CDF collaboration has presented preliminary results from the Tevatron collider. They search over the dimuon mass ranges 3.2-3.5 and 3.8-4.4 GeV, and again extrapolate to the full dimuon mass range to get upper limits on the short distance contribution of $3.5 \times 10^{-5}$ and $5.3\times 10^{-5}$ for the exclusive channels $B^0 \to K^{*0}\mu^+\mu^-$ and $B^- \to K^-\mu^+\mu^-$, respectively [51] .

In contrast to the hadron collider experiments CLEO has searched for all dilepton masses except for the ranges 2.9-3.2 and 3.5-3.8 GeV where the $\psi^{(\prime)}$ contributions dominate [52]. The analysis uses standard methods to reconstruct exclusive $B$ meson decays from a candidate $K$ or $K^*$ meson and a pair of identified leptons. Some typical plots of the beam-constrained mass distributions are shown in Figure 8. In an $\Upsilon(4S)$

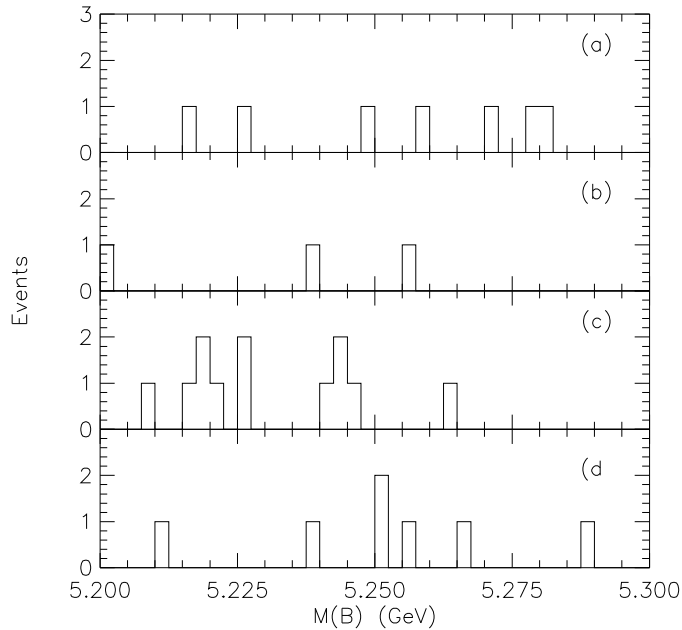

Figure 8: Beam constrained mass distributions from CLEO for $B \to K^{(*)}\ell^+\ell^-$: (a) $K^+e^+e^-$ (b) $K^+\mu^+\mu^-$ (c) $K^{*0}e^+e^-$ (d) $K^{*0}\mu^+\mu^-$

data sample of 2.0 $fb^{-1}$ the background is less than one event in the signal region for each of the exclusive channels. The residual background is half from the continuum and half from $B\bar{B}$ events where both $B$ mesons decay semileptonically. Table 2 summarizes the preliminary upper limits from CLEO for the exclusive channels $B \to K^{(*)}\mu^+\mu^-$ and $B \to K^{(*)}e^+e^-$. The rate for the decays to electron pairs is predicted to be larger than that to muon pairs due to the contribution from low mass pairs below the dimuon mass threshold. In some cases the limits from CLEO are close to the theoretical expectations. In the future significant increases in statistics at both hadron colliders and at $\Upsilon(4S)$ machines are expected to lead to the observation of $b \to s\ell^+\ell^-$. Eventually there should be enough statistics to measure the dilepton mass distribution, and other kinematic variables characterizing the three-body final state. Of particular interest is the forward-backward asymmetry of the lepton pair, since this is expected to be large in the Standard Model, and may be rather sensitive to non-Standard Model physics [26, 54].

Table 3: Results of $b \to s\ell^+\ell^-$ searches at CLEO.

| $B$ Decay Mode | Candidate Events | Detection Efficiency | 90% C.L. Upper Limit | Standard Model Prediction[53] |
|---|---|---|---|---|
| $K^+e^+e^-$ | 2 | 24.4% | $12.0 \times 10^{-6}$ | $0.6 \times 10^{-6}$ |
| $K^+\mu^+\mu^-$ | 0 | 15.1% | $9.0 \times 10^{-6}$ | $0.6 \times 10^{-6}$ |
| $K^{*0}e^+e^-$ | 0 | 9.8% | $16.0 \times 10^{-6}$ | $5.6 \times 10^{-6}$ |
| $K^{*0}\mu^+\mu^-$ | 0 | 5.0% | $31.0 \times 10^{-6}$ | $2.9 \times 10^{-6}$ |

## VI. Rare Hadronic Decays

Rare hadronic $B$ decays are described by a combination of a $b \to u$ spectator diagram (Figure 1(a)), and a gluonic penguin diagram (Figure 1(d)). Decay modes such as $\pi^+\pi^-$ and $\pi^\pm\rho^\mp$ are expected to be described mainly by the spectator diagram with a $\pi^-$ or $\rho^-$ being produced by the $W^-$. A small contribution from a $b \to d$ penguin diagram is also expected in these modes. Decay modes such as $K^-\pi^+$ and $K^{*-}\pi^+$ are expected to result mainly from a $b \to s$ penguin diagram, with a small contribution coming from the Cabibbo-suppressed spectator diagram where the $W^-$ produces a $K^{(*)-}$. There are also a few decay modes, such as $K^0\pi^0$ and $K^{(*)}\phi$ that are described only by a gluonic penguin diagram, and a few modes such as $\pi^-\pi^0$ that are described only by a spectator diagram. In modes where both penguin and spectator diagrams are significant direct CP violation can occur, as will be discussed in Section VII.

To establish the relative importance of the penguin and spectator amplitudes it is necessary to study a large number of decay modes. The first evidence for hadronic $B$ meson decays to final states without charmed mesons came from the CLEO observation of a signal in the sum of the two decay modes $B^0 \to K^+\pi^-$ and $B^0 \to \pi^+\pi^-$ [55]. We will refer to this sum as $B^0 \to h^+\pi^-$. Since this publication the CLEO data sample has increased by almost a factor of two, and a number of other charmless hadronic decay modes have been studied. We also note that there are a few candidate events for $B^0 \to h^+\pi^-$ or $B_s \to h^+K^-$ from the DELPHI and ALEPH experiments at LEP [56].

### A) Decays to $K\pi$ and $\pi\pi$ Final States

The signature for a $B^0$ decay to two charged tracks is a particularily simple one. At the $\Upsilon(4S)$ the $B$ is almost at rest, and the tracks are back-to-back with momenta of about 2.6 GeV. CLEO observed a signal for such events in 1.4 $fb^{-1}$ of $\Upsilon(4S)$ data [55]. Here we discuss new results from a larger data sample of 2.4 $fb^{-1}$ [57].

The same two kinematical variables are used as in the $B \to K^*\gamma$ analysis (Section

III(A)), i.e. the energy sum of the two tracks relative to the beam energy ($\Delta E$), which has an r.m.s. resolution of 25 MeV, and the beam constrained invariant mass ($m_B$), which has an r.m.s. resolution of 2.6 MeV. Particle identification uses $dE/dx$ information from the main tracking chamber. Separation between the $K^+\pi^-$ and $\pi^+\pi^-$ hypotheses comes from the $dE/dx$ information (1.8$\sigma$, where $\sigma$ is the rms resolution), and from the difference in $\Delta E$ (1.7$\sigma$). The overall separation of 2.5$\sigma$ is rather marginal, and is expected to be much better in future detectors [48].

The background is due to continuum production of two light quark jets. From studies of off-resonance data samples it is known that a cut on the thrust axis, $\cos\theta_T$, discussed in Section III(A), is most effective against this background. Requiring $\cos\theta_T < 0.7$ removes 95% of the background and only 35% of the signal. There is some additional discrimination from the energy distribution of the rest of the event, the direction defined by the axis of the two tracks, and the direction of the $B$ meson. This information is combined into one variable ($\mathcal{F}$), using a linear Fisher discriminant technique [58].

The final signal yields are obtained from a likelihood fit to the four variables, $\Delta E$, $m_B$, $\mathcal{F}$ and $dE/dx$, using an event sample containing the signal region and a large sideband in $\Delta E$ and $m_B$ from which the background is determined. In the first version of this fit all three signal hypotheses are allowed, $\pi^+\pi^-$, $K^+\pi^-$ and $K^+K^-$. It is found that the best fit has zero yield for a $K^+K^-$ signal. This is expected since this decay mode cannot occur via a penguin or spectator diagram. After setting an upper limit of $4.0 \times 10^{-6}$ (90% C.L.) on $B^0 \to K^+K^-$, a second fit is done in which only the first two signal hypotheses are included. The results of this fit projected onto the $m_B$ and $\Delta E$ axes, are shown as the solid and dotted lines in Figure 9. The event histograms result from an event-counting analysis that will be described in section VI(B).

The statistical significance of the signal yield in Figure 9 is determined from the probability that the fitted background fluctuates up to the combined yield of signal plus background. Although this is determined by Poisson statistics for these small event samples, it is conventional to quote the probability in the equivalent number of $\sigma$ of a Gaussian distribution. With this definition, the combined significance of the two signal modes is quoted as 5$\sigma$, with each individual mode having a significance of about 2.5$\sigma$. These results are interpreted as an observation of the sum of the two decays, but not yet as a significant result for either of the individual channels.

The combined branching ratio for $B^0 \to K^+\pi^-$ and $B^0 \to \pi^+\pi^-$ is measured to be $(1.8 \pm 0.6 \pm 0.2) \times 10^{-5}$. The signal yields and the upper limits on the individual branching fractions are given in Table 4. These results are consistent with the theoretical predictions given in the last column of Table 4.

CLEO has made a similar analysis of the decay modes $B^+ \to h^+\pi^0$. Here the continuum background is larger and the $K/\pi$ separation is weaker, since the presence of a $\pi^0$ leads to a $\Delta E$ resolution of 50 MeV. There are also results for the decay modes $B^0 \to \pi^0\pi^0$, $B^+ \to K^0\pi^+$ and $B^0 \to K^0\pi^0$. In these three cases only one signal hypothesis is assumed and no $dE/dx$ information is used in the likelihood fit. The mass distributions for these modes are shown in Figure 10, and the results of the likelihood fits are summarized in Table 4. It is found that no individual mode has a significance greater than $3\sigma$, although most of them are fitted with a small positive yield. With the exception of $\pi^0\pi^0$ the other modes in Table 4 are expected to be observed with branching fractions comparable to or just below the $h^+\pi^-$ channels, so it is likely that many of these modes will be observed in the near future.

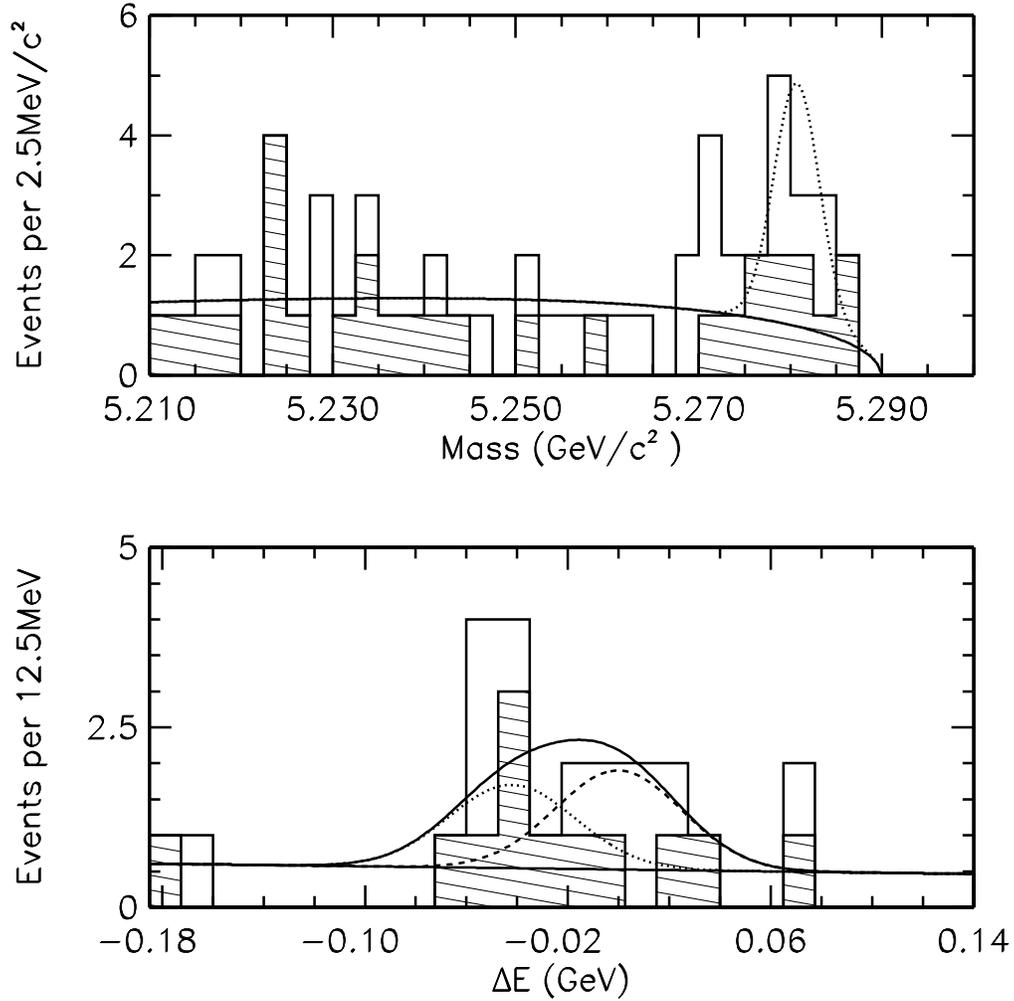

Figure 9: Projections of the $B^0 \to h^+\pi^-$ candidates onto the $m_B$ and $\Delta E$ variables. The lines show the result of the likelihood fit. In the upper plot the solid line is the fitted background, and the dotted line is the fitted signal. In the lower plot the lower solid line is the background, the dashed line is the fitted $B^0 \to \pi^+\pi^-$ signal the dotted line is the fitted $B^0 \to K^+\pi^-$ signal and the upper curve is the sum of all three contributions. Shaded events are identitifed as $K^+\pi^-$, unshaded as $\pi^+\pi^-$.

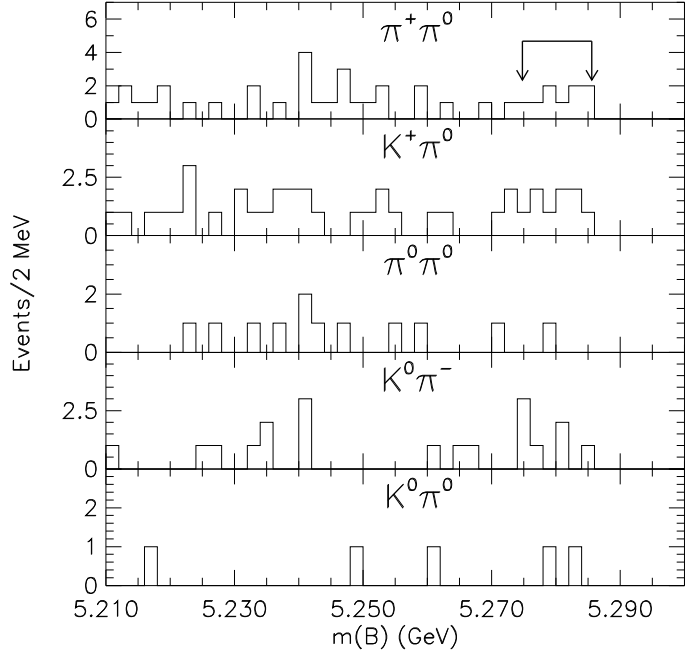

Figure 10: Beam constrained mass distributions for rare hadronic decays to pseudoscalar mesons. The arrows indicate the signal region.

Table 4: Results of CLEO II Searches for Rare Hadronic $B$ Decays to Two Pseudoscalar Mesons.

| $B$ Decay Mode | Signal Yield (Likelihood Fit) | Fitted Background | B.R. (90% C.L.) $\times 10^{-5}$ | Predictions $\times 10^{-5}$ [22, 59] |
|---|---|---|---|---|
| $\pi^+\pi^-$ | $9.4^{+4.9}_{-4.1}$ | $5.8\pm 0.3$ | $< 2.0$ | 1.0-2.6 |
| $K^+\pi^-$ | $7.9^{+4.5}_{-3.6}$ |  | $< 1.7$ | 1.0-2.0 |
| $\pi^+\pi^0$ | $5.0^{+4.2}_{-3.2}$ | $12.6\pm 0.5$ | $< 1.7$ | 0.6-2.1 |
| $K^+\pi^0$ | $4.9^{+3.6}_{-2.8}$ |  | $< 1.4$ | 0.3-1.3 |
| $\pi^0\pi^0$ | $1.2^{+1.7}_{-0.9}$ | $2.1\pm 0.2$ | $< 0.9$ | 0.03-0.10 |
| $K^0\pi^+$ | $5.2^{+3.5}_{-2.8}$ | $1.6\pm 0.1$ | $< 4.8$ | 1.1-1.2 |
| $K^0\pi^0$ | $2.3^{+2.2}_{-1.5}$ | $0.7\pm 0.1$ | $< 4.0$ | 0.5-0.8 |

## B) Decays to Vector and Pseudoscalar Mesons

In this section we discuss searches for the decays $B \to \pi\rho$, $B \to K\rho$, $B \to K^*\pi$ and $B \to K^{(*)}\phi$. The final states for these decays contain three or more particles, up to two of which may be $\pi^0$s, and one of which may be a $K_s$. The $\rho$, $K^*$ and $\phi$ final states are selected by a cut of one natural width about the resonance mass. The $dE/dx$ information from the main tracking chamber is used to select the most probable decay mode in cases where this is ambiguous.

In contrast to the previous section, CLEO has used a simple event-counting analysis to search for these modes rather than a full likelihood fit. Cuts are made on $\cos\theta_T$, $\mathcal{F}$ and $dE/dx$. In addition for decays to a vector and a pseudoscalar meson the decay helicity angle, $\theta_H$, is defined as the angle in the vector meson rest frame between the direction of the $B$ meson and one of the decay products of the vector meson. Since signal events have a $\cos^2\theta_H$ distribution, a $|\cos\theta_H| > 0.5$ cut can be used to suppress continuum background. After all these cuts have been made, the event yield in the signal region in the $m_B - \Delta E$ plane is compared to the yield expected from an extrapolation of a large two-dimensional sideband region.

Table 5 summarizes the results from 2.4 $fb^{-1}$ of $\Upsilon(4S)$ data. In most cases there are few events in the signal region. In the $K^0\rho$ and $K^{(*)}\phi$ channels there are also few events in the sideband region, and we do not quote an estimated background number because of the difficulty in extrapolating the yield from such small statistics. There are no significant signals in any of the decay modes in Table 5, although $\pi^\pm\rho^\mp$ and $K^{*+}\pi^-$ do have more events in the signal region than expected purely from background. The upper limits on $B^0 \to \pi^\pm\rho^\mp$ and $B^+ \to K^+\phi$ are close to the theoretical predictions.

## VII. CP violation in Rare Decays

In the Standard Model CP violation arises from a complex phase in the CKM matrix, which relates the mass eigenstates to the weak eigenstates. This is an inevitable consequence of having three families of quarks. In general, if we have two interfering amplitudes, we can write each of them as a product of a strong decay amplitude and a weak decay amplitude

$$\mathcal{A} = a_s e^{i\theta_s} a_w e^{i\theta_w}$$

Table 5: Results of CLEO II Searches for Rare Hadronic $B$ Decays to Final States with Vector Mesons.

| $B$ Decay Mode | Event Yield (Signal Region) | Estimated Background | B.R.(90% C.L.) $\times 10^{-5}$ | Prediction $\times 10^{-5}$ [22, 59] |
|---|---|---|---|---|
| $\pi^\pm \rho^\mp$ | 7 | 2.9±0.7 | < 8.8 | 1.9-8.8 |
| $\pi^0 \rho^0$ | 1 | 1.8±0.6 | < 2.4 | 0.07-0.23 |
| $\pi^+ \rho^0$ | 4 | 2.3±0.3 | < 4.3 | 0.0-1.4 |
| $\pi^0 \rho^+$ | 8 | 5.5±1.2 | < 7.7 | 1.5-3.9 |
| $K^{*+} \pi^-$ | 3 | 0.7±0.2 | < 7.2 | 0.1-1.9 |
| $K^{*0} \pi^0$ | 0 | 1.1±0.3 | < 2.8 | 0.3-0.5 |
| $K^{*+} \pi^0$ | 4 | 1.9±0.7 | < 9.9 | 0.1-0.9 |
| $K^{*0} \pi^+$ | 2 | 1.0±0.6 | < 4.1 | 0.6-0.9 |
| $K^+ \rho^-$ | 2 | 2.0±0.4 | < 3.5 | 0.00-0.20 |
| $K^0 \rho^0$ | 0 | | < 3.9 | 0.01-0.04 |
| $K^+ \rho^0$ | 1 | 3.8±0.2 | < 1.9 | 0.01-0.06 |
| $K^0 \rho^+$ | 0 | | < 4.8 | 0.00-0.03 |
| $K^0 \phi$ | 1 | | < 8.8 | 0.1-1.3 |
| $K^{*0} \phi$ | 2 | | < 4.3 | 0.0-3.1 |
| $K^+ \phi$ | 0 | | < 1.2 | 0.1-1.5 |
| $K^{*+} \phi$ | 1 | | < 7.0 | 0.0-3.1 |

$$\mathcal{B} = b_s e^{i\delta_s} b_w e^{i\delta_w}. \qquad (13)$$

Applying the CP operators to these amplitudes results in

$$\overline{\mathcal{A}} = a_s e^{i\theta_s} a_w e^{-i\theta_w}$$
$$\overline{\mathcal{B}} = b_s e^{i\delta_s} b_w e^{-i\delta_w}. \qquad (14)$$

Note that the weak phase has changed sign, while the strong phase has not. The rate difference, which may exhibit CP violation is

$$\Gamma - \overline{\Gamma} = |\mathcal{A} + \mathcal{B}|^2 - |\overline{\mathcal{A}} + \overline{\mathcal{B}}|^2 = 2 a_s a_w b_s b_w \sin(\delta_s - \theta_s) \sin(\delta_w - \theta_w). \qquad (15)$$

If two distinct weak decays processes are possible which go via CKM elements with a phase difference, then $\sin(\delta_w - \theta_w) \neq 0$. Guaranteeing a strong phase shift, however, is not possible. In fact, the theory of strong decays lacks sufficient power to be able to accurately predict the magnitude of such phase differences, or their sign relative to the weak phase.

## A) CP violation in $B^\pm$ Decays to Two Pseudoscalars

Direct CP violation can occur in charged $B$ meson decays due to interference between any two diagrams with different weak and strong phases. In decays to two pseudoscalar mesons the penguin and tree diagrams can give rise to integral rate asymmetries such as:

$$\Delta(K\pi^0) = \frac{\Gamma(B^- \to K^-\pi^0) - \Gamma(B^+ \to K^+\pi^0)}{\Gamma(B^- \to K^-\pi^0) + \Gamma(B^+ \to K^+\pi^0)} \tag{16}$$

that are manifestly CP violating.

There have been several suggestions for measurements of decay rates of $B^\pm$ mesons that could be used to determine the CKM phase $\sin\gamma$ [60, 61]. Although the discussion of the derivation of $\sin\gamma$ is complicated, we would like to present the arguments for comparing the rates of $B^\pm$ decays to two pseudoscalar mesons in order to extract $\sin\gamma$ [60], since it is likely that future experiments will measure these decay rates [48].

The final state $K^-\pi^0$ has an amplitude ($\mathcal{T}_s$) from the tree diagram in Figure 11(a) if the $W^-$ materializes as a $K^-$, an amplitude ($\mathcal{C}_s$) from the "color suppressed" diagram 11(b), and an amplitude ($\mathcal{P}_s$) from the penguin diagram in Figure 11(c). For the final state $\pi^-\pi^0$ there are analogous tree ($\mathcal{T}$) and color suppressed ($\mathcal{C}$) amplitudes where the $s$ quark in these diagrams is replaced by a $d$ quark. However there is no analogous penguin amplitude because the gluon can form $d\bar{d}$ as well as $u\bar{u}$ and these amplitudes cancel. This is the same as the statement that the I=$\frac{3}{2}$ $\pi^+\pi^0$ final state cannot be made with a $\Delta$I=$\frac{1}{2}$ penguin amplitude. For the $K^0\pi^-$ final state there is only one contribution from the penguin diagram shown in Figure 11(d).

Assuming SU(3) symmetry, the strange amplitudes are related to the non-strange amplitudes by:

$$\frac{\mathcal{T}_s}{\mathcal{T}} = \frac{\mathcal{C}_s}{\mathcal{C}} = r = \frac{V_{us}}{V_{ud}}\frac{f_K}{f_\pi} = 0.28 \tag{17}$$

The amplitudes can be summarized as

$$\mathcal{A}(B^- \to \pi^-\pi^0) = -\frac{1}{\sqrt{2}}(\mathcal{T} + \mathcal{C}) \tag{18}$$

$$\mathcal{A}(B^- \to K^0\pi^-) = \mathcal{P}_s \tag{19}$$

$$\mathcal{A}(B^- \to K^-\pi^0) = -\frac{1}{\sqrt{2}}(\mathcal{T}_s + \mathcal{C}_s + \mathcal{P}_s) \tag{20}$$

These amplitudes can be related by

$$\sqrt{2}\mathcal{A}(B^- \to K^-\pi^0) + \mathcal{A}(B^- \to K^0\pi^-) = r\sqrt{2}\mathcal{A}(B^- \to \pi^-\pi^0). \tag{21}$$

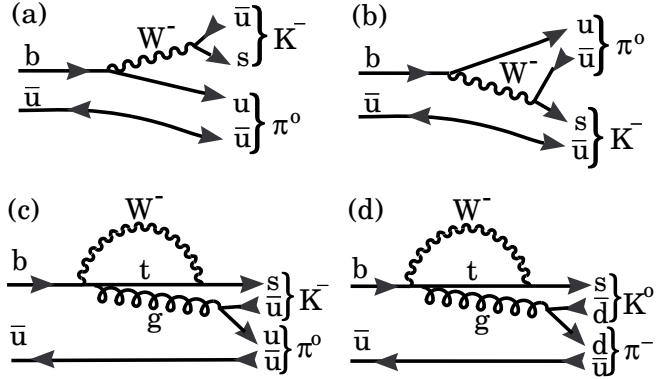

Figure 11: Amplitudes contributing to $B^-$ decays to two pseudoscalars. (a) tree diagram for $K^-\pi^0$ (b) color-suppressed diagram for $K^-\pi^0$ (c) penguin diagram for $K^-\pi^0$ (d) penguin diagram for $K^0\pi^-$. The $\pi^-\pi^0$ final state goes through (a) and (b), with the $W^- \to \bar{u}d$ rather than $\bar{u}s$.

Amplitude triangles can be constructed from this relationship and its complex conjugate for $B^+$ decays (Figure 12).

Since $\mathcal{A}(B^- \to K^0\pi^-)$ involves only one penguin diagram, it is equal to $\mathcal{A}(B^+ \to K^0\pi^+)$. On the other hand,

$$r\sqrt{2}\mathcal{A}(B^- \to \pi^-\pi^0) = a_T e^{i\delta_T} e^{-i\gamma} \tag{22}$$

$$r\sqrt{2}\mathcal{A}(B^+ \to \pi^+\pi^0) = a_T e^{i\delta_T} e^{i\gamma}, \tag{23}$$

where $\gamma = \arg(V_{ub}^* V_{us})$. Note, that the rates of these two processes are equal since they involve a single weak phase and a single strong phase; however, there is a difference in phase of $2\gamma$ between them. In the case of $K^-\pi^0$ and $K^+\pi^0$ the penguin and tree contributions interfere and there is a net weak phase shift of $\arg(V_{ub} V_{us}^* V_{tb}^* V_{ts})$. The strong phase shift, $\delta = \delta_T - \delta_P$, is also important in determining the actual rate asymmetry, $\Delta(K\pi^0)$, which has been estimated to be a few percent by several authors [62]. CP violation would be explicit if a measured rate difference between $K^-\pi^0$ and $K^+\pi^0$ existed, but this requires a strong phase shift as well as the weak phase shift. It has been argued that by constructing amplitude triangles as shown in Figure 12, the angle $\gamma$ can be determined with a twofold ambiguity. Note that if $\delta$ is zero $\gamma$ can be derived unambiguously even though there is no explicit CP violation in $B^\pm \to K^\pm\pi^0$ [60].

This procedure is only valid if there are no additional contributions other than those discussed above. For example the $c$ and $u$ quark penguin loops could generate significant asymmetries with different phases from the $t$ quark penguin. There are several experimental checks that other diagrams can be neglected. The decays $B^- \to K^0\pi^-$ and $B^+ \to K^0\pi^+$ should have equal rates, as should $B^- \to \pi^-\pi^0$ and $B^+ \to \pi^+\pi^0$. If non-equal rates were observed in these decays this would also be an observation of direct CP violation, but it would not be possible to extract the angle $\gamma$ from the triangle relation any more. Another check comes from $B^0 \to K^+K^-$, since this cannot be produced by the tree and penguin diagrams of Figure 11. It is important that this final state not be observed at a small fraction of the other decays.

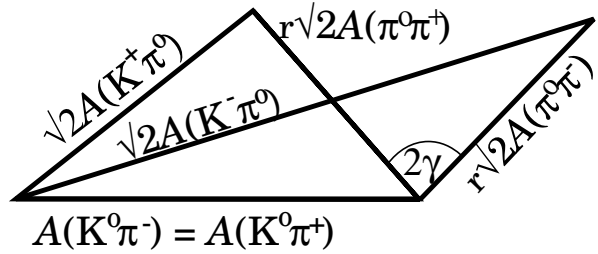

Figure 12: Amplitude triangles relating $B^{\pm} \to K^{\pm}\pi^0$, $B^{\pm} \to \pi^{\pm}\pi^0$ and $B^{\pm} \to K^0\pi^{\pm}$.

Deshpande and He [63] have pointed out that there is another class of penguin diagrams where the gluon in Figure 11(c) and (d) is replaced by a $\gamma$, $Z$, or a box diagram. Gluonic penguins couple equally to $u\bar{u}$ and $d\bar{d}$, whereas the $\gamma$ and $Z$ penguins couple differently to $u$ and $d$ quarks. These electroweak penguin contributions are expected to be small in $B^{\pm} \to \pi^{\pm}\pi^0$, but they could be as large as the tree level contributions in $B^{\pm} \to K^{\pm}\pi^0$. If this is the case, the method we have described above cannot be used to extract $\gamma$ without information from other decay modes.

There have been two suggestions on how to extract $\sin\gamma$ allowing for possible electroweak penguin contributions. Gronau et al. suggest forming an amplitude quadrangle including the additional decay mode $B_s \to \eta\pi^0$ [64]. However, measuring this rare $B_s$ decay appears to be extremely difficult. Deshpande and He suggest in a recent preprint the construction of additional amplitude triangles using the decays $B^- \to K^-\eta^{(\prime)}$ and $B^+ \to K^+\eta^{(\prime)}$ [65]. The octet part of the $\eta/\eta'$ system is defined as $\eta_8$:

$$\mathcal{A}(K^-\eta_8) = \mathcal{A}(K^-\eta)\cos\theta + \mathcal{A}(K^-\eta')\sin\theta, \qquad (24)$$

where $\theta$ is the $\eta - \eta'$ mixing angle of about $20^\circ$ [6]. Two new amplitude triangles can be constructed:

$$\sqrt{2}\mathcal{A}(K^-\pi^0) - 2\mathcal{A}(\bar{K}^0\pi^-) = \sqrt{6}\mathcal{A}(K^-\eta_8) \qquad (25)$$

$$\sqrt{2}\mathcal{A}(K^+\pi^0) - 2\mathcal{A}(K^0\pi^+) = \sqrt{6}\mathcal{A}(K^+\eta_8). \qquad (26)$$

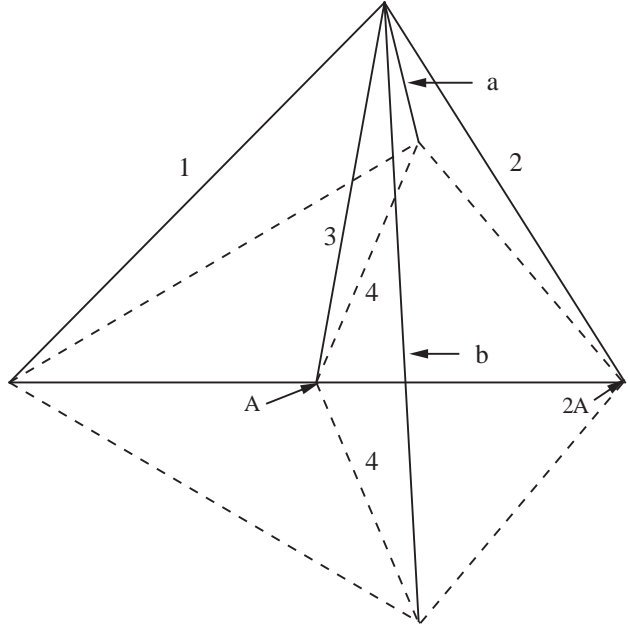

Figure 13: Amplitude triangles and two solutions **a** and **b** for the magnitude of $\sin\gamma$. The amplitude A represents $\mathcal{A}(K^0\pi^-)$. The amplitude 2A forms a triangle with the amplitudes $\sqrt{2}\mathcal{A}(K^-\pi^0)$ (line 1) and $\sqrt{6}\mathcal{A}(K^-\eta_8)$ (line 2). This triangle determines the amplitude $\mathcal{B}$ (line 3). The dotted lines show the equivalent construction for the $B^+$ decays, leading to the amplitude $\overline{\mathcal{B}}$ (line 4). There is a two fold ambiguity in the angle between 3 and 4 due to the possibility of flipping the $B^+$ triangle with respect to the $B^-$ triangle. This leads to the two results for $\mathcal{B} - \overline{\mathcal{B}}$ shown as **a** and **b**.

It is then convenient to construct the amplitude combinations

$$\mathcal{B} = \sqrt{2}\mathcal{A}(K^-\pi^0) - \mathcal{A}(\bar{K}^0\pi^-) \tag{27}$$

$$\overline{\mathcal{B}} = \sqrt{2}\mathcal{A}(K^+\pi^0) - \mathcal{A}(K^0\pi^+). \tag{28}$$

The complete amplitude construction is shown in Figure 13. The solid lines show the amplitude triangle for the $B^-$ decays (equation (25)), while the dotted lines show the $B^+$ amplitudes (equation (26)). From these triangles $\mathcal{B}$ (equation (27)), and $\overline{\mathcal{B}}$ (equation (28)), are constructed. The difference $\mathcal{B} - \overline{\mathcal{B}}$ (equation (29)), has two possible solutions **a** and **b**, which are related to the amplitude for $B^- \to \pi^-\pi^0$ and to $\sin\gamma$ by:

$$\mathcal{B} - \overline{\mathcal{B}} = -i2\sqrt{2}e^{i\delta}\left|\frac{V_{us}}{V_{ud}}\right|\left|\mathcal{A}(\pi^-\pi^0)\right|\sin\gamma, \tag{29}$$

where $\delta$ is a strong phase shift. From the measured difference from the triangle construction **a** or **b**, and the measured rate for $B^- \to \pi^-\pi^0$ the angle $\gamma$ can be determined with a twofold ambiguity.

The calculation of Deshpande and He assumes that the couplings of kaons, pions and etas are related by SU(3), and that the decay amplitudes can be factorized. In equation (29) the amplitude $\mathcal{A}(\pi^-\pi^0)$ should be multiplied by a factor $(f_K/f_\pi)$, and in equations (25) and (26) the amplitude $\mathcal{A}(K^-\eta_8)$ by a factor $(f_\pi/f_\eta)$. Although there is some uncertainty in these theoretical assumptions, the success of factorization in explaining $B$ decays to exclusive final states with a $D^{*+}$ and a light hadron has been encouraging [66].

The angle $\gamma$ can also be determined using measured rates in charged $B$ decays to $D^0 K$ final states. The method proposed by Gronau and Wyler [61] uses the three related decay modes $B^- \to D^0 K^-$, $B^- \to \bar{D}^0 K^-$, $B^- \to D_{CP} K^-$, where $D_{CP}$ indicates that the $D^0$ decays into a CP eigenstate, and the corresponding modes for $B^+$. The decay $B^- \to D^0 K^-$ is a Cabibbo suppressed version of $B^- \to D^0 \pi^-$, while the decay $B^- \to \bar{D}^0 K^-$ is a color suppressed $b \to u$ transition where the virtual $W^-$ transforms itself into a $\bar{c}s$ pair. Interference is possible between these two decays modes if the $D^0$ decays into a CP eigenstate. Examples of such final states include $K^+K^-$, $K_s\pi^0$, and $K_s\eta$. To simplify the discussion only states that are in specific angular momentum configurations are used so that their CP is defined as +1 or -1; these states are usually denoted as $D_1^0$ and $D_2^0$. We have

$$D_1^0 = \frac{1}{\sqrt{2}}\left[D^0 + \bar{D}^0\right], \quad \text{and} \quad D_2^0 = \frac{1}{\sqrt{2}}\left[D^0 - \bar{D}^0\right].$$

The amplitudes for the three $B^-$ decays modes are related by:

$$\sqrt{2}\mathrm{A}_1^-(B^- \to D_1^0 K^-) = \mathrm{A}(B^- \to \bar{D}^0 K^-) + \overline{\mathrm{A}}(B^- \to D^0 K^-).$$

Denoting the hadronic phase as $\delta$, gives

$$\sqrt{2}\mathrm{A}_1^-(B^- \to D_1^0 K^-) = |\mathrm{A}|e^{i(\phi_s+\delta)} + \overline{\mathrm{A}}e^{i\delta}.$$

The decays to $D_1^0$ need not be equal for $B^+$ and $B^-$, and an asymmetry in them is a manifest demonstration of CP violation. A triangle construction serves to determine $\sin\gamma$. For more details see [61, 67].

## B) CP violation in $B^0 \to \pi^+\pi^-$ due to mixing

The final state $\pi^+\pi^-$ is one of the simplest in $B$ decay, since there are only two pseudoscalar particles in the final state, and there is no spin or angular momentum to consider. This final state can be reached from either a $B^0$ or a $\bar{B}^0$, and the s-wave production of the two spinless particles means that this is a CP eigenstate. The two interfering amplitudes necessary for CP violation are provided by the direct $B^0$ decay, and the indirect decay following $\bar{B}^0 - B^0$ mixing. When mixing provides the second amplitude for a decay to a CP eigenstate, the strong phase shift disappears from the equation relating the measured CP asymmetry to the CKM angles.

To measure CP violation using mixing we need to make use of the correlated production of $B^0\bar{B}^0$ pairs. This is done by tagging the number of $\pi^+\pi^-$ events produced with the other $B$ decaying as a $B^0$, as opposed to a $\bar{B}^0$. An example of a suitable tag is lepton flavor. These numbers are time dependent functions of $T = t - t'$ (in units of mean lifetime), where $t$ is the decay time of the $\pi^+\pi^-$ and $t'$ is the decay time of the other $B$:

$$\mathcal{R}(T) \propto e^{-|T|}(1 - \sin\frac{\Delta m}{\Gamma}T \ \sin 2\alpha) \quad (\text{for } \pi^+\pi^-, B^0), \tag{30}$$

$$\overline{\mathcal{R}}(T) \propto e^{-|T|}(1 + \sin\frac{\Delta m}{\Gamma}T \ \sin 2\alpha) \quad (\text{for } \pi^+\pi^-, \bar{B}^0). \tag{31}$$

It can be shown that $\overline{\mathcal{R}}(T) = \mathcal{R}(-T)$. The CP asymmetry is

$$\mathcal{A}_T(\pi^+\pi^-) = \frac{\mathcal{R}(T) - \overline{\mathcal{R}}(T)}{\mathcal{R}(T) + \overline{\mathcal{R}}(T)} = \sin\frac{\Delta m}{\Gamma}T \ \sin 2\alpha \tag{32}$$

We can measure a time independent asymmetry by integrating over $t-t'$. However, if the C parity of the intial $B^0\bar{B}^0$ pair is -1, as is the case in $\Upsilon(4S)$ decay, the time integrated rate is zero. This is not the case in hadron colliders, or if the initial state is $B^{0*}\bar{B}^0$.

Measuring CP violation in the $\pi^+\pi^-$ decay determines sin$2\alpha$, where

$$\alpha = \arg(V_{ud}V_{ub}^*/V_{td}V_{tb}^*). \tag{33}$$

However, there is a problem due to the presence of a decay amplitude related to the penguin diagram shown in Figure 1(d). A $\pi^+\pi^-$ final state is produced when the $t$ quark in the loop couples to a $d$ quark rather than an $s$ quark. Although this is suppressed it could turn out to have a significant effect on the CP asymmetry measurement. Gronau and London [68] have shown how this contribution can be isolated by measuring the rates for the $\pi^+\pi^0$ and $\pi^0\pi^0$ processes. Equations (30) and (31) are modified as follows:

$$\mathcal{R}(T) \propto e^{-|T|}\Big(\frac{1+|\xi|^2}{2} + \frac{1-|\xi|^2}{2}\cos\frac{\Delta m}{\Gamma}T + \mathcal{I}m\xi \ \sin\frac{\Delta m}{\Gamma}|T|\Big) \tag{34}$$

$$\overline{\mathcal{R}}(T) \propto e^{-|T|}\Big(\frac{1+|\xi|^2}{2} - \frac{1-|\xi|^2}{2}\cos\frac{\Delta m}{\Gamma}T - \mathcal{I}m\xi \ \sin\frac{\Delta m}{\Gamma}|T|\Big), \tag{35}$$

where $\xi$ is a parameter related to $\alpha$. If only the direct and the mixing amplitude are present $|\xi| = 1$, and we are left with equations (30) and (31). We can look for the presence of the cosine term experimentally by forming a new time dependent asymmetry:

$$\mathcal{A}_{|T|}(\pi^+\pi^-) = \frac{\overline{\mathcal{R}}(|T|) + \overline{\mathcal{R}}(-|T|) - \mathcal{R}(|T|) - \mathcal{R}(-|T|)}{\overline{\mathcal{R}}(|T|) + \overline{\mathcal{R}}(-|T|) + \mathcal{R}(|T|) + \mathcal{R}(-|T|)} = \frac{1-|\xi|^2}{1+|\xi|^2}\cos\frac{\Delta m}{\Gamma}|T|. \tag{36}$$

The asymmetry $\mathcal{A}_{|T|}$ leads to a non-vanishing asymmetry, $\Delta$, in the time integrated rates, even at the $\Upsilon(4S)$:

$$\Delta(\pi^+\pi^-) = \frac{1-|\xi|^2}{1+|\xi|^2}\frac{1}{1+(\frac{\Delta m}{\Gamma})^2} \qquad (37)$$

If it turns out that the $\pi^0\pi^0$ rate is comparable to $\pi^+\pi^-$, the penguin amplitude is important but it should be possible to measure it. If it is small enough to be difficult to measure, then the penguin amplitude is likely to be unimportant and $\alpha$ comes out simply.

Another method for extracting the size of the penguin-tree interference is to compare the integral rates for $B^0 \to \pi^+\pi^-$ and $B^0 \to K^+\pi^-$ with those for $\bar{B}^0 \to \pi^-\pi^+$ and $\bar{B}^0 \to K^-\pi^+$ [69]. A recent preprint by Deshpande and He [70] shows that the assumption of SU(3) symmetry leads to the following simple relationship between the time integrated rate asymmetries for these decays:

$$\Delta(\pi^+\pi^-) \approx -\frac{f_\pi^2}{f_K^2}\Delta(K^+\pi^-) \qquad (38)$$

and that this result can be used to correct the measurements of the time-dependent CP asymmetry in $\pi^+\pi^-$ for the effect of the penguin amplitude.

## C) CP violation in radiative penguin decays

A measurement of the ratio of $b \to d\gamma$ to $b \to s\gamma$ determines $|V_{td}|/|V_{ts}|$ if the amplitudes are described by the $t$ quark loop. As discussed in Section V(A) there may be other amplitudes that are significant in $b \to d\gamma$. If this is the case then these additional amplitudes can give rise to direct CP violation in these decays.

One source of CP violation is the presence of three different loop diagrams involving the $u$, $c$ or $t$ quarks. Gluon exchange provides the necessary strong phase shifts between these loop diagrams. Naively, one would expect that the $u$ and $c$ diagrams would be highly suppressed due to the relatively small quark masses. However both Deshpande et al. [45] and Soares [71] find significant $u$ and $c$ loop contributions in $b \to d\gamma$. Soares has explicitly calculated the amount of CP violation from these sources and finds asymmetries of $(2-8) \times 10^{-3}$ for $b \to s\gamma$ and a significantly higher value of $(2-30) \times 10^{-2}$ for $b \to d\gamma$. For the exclusive decay modes Greub et al. [72] estimate that the CP asymmetries are 1% for $B \to K^*(890)\gamma$ and 15% for $B \to \rho\gamma$.

There are other diagrams that can provide a source of CP violation. Of particular interest is the suggestion that non Standard Model contributions to $b \to s\gamma$ can lead to large CP asymmetries, since the asymmetry predicted by the Standard Model is rather small. As an example of such possibilities we mention the paper by Wolfenstein and Wu [73] that calculates the expected level of CP violation in a two Higgs doublet model. Here the charged Higgs is present in the loop instead of the $W^-$, and there is an arbitrary phase factor associated with the two Higgs doublets. Depending on this phase the asymmetry in $b \to s\gamma$ could be anywhere in the range 0-10%. If a CP asymmetry larger than 1% were observed in $b \to s\gamma$, it would provide strong evidence for physics beyond the Standard Model.

# VIII. Conclusions

Loop diagrams were first discovered in the mixing amplitude for neutral kaon decays. Mixing has also been observed in the neutral $B$ mesons [Figure 1(b)]. CLEO has established the existence of the radiative penguin decay $b \to s\gamma$ [Figure 1(c)], by measuring the exclusive branching fraction for $B \to K^*\gamma$ to be $(4.5 \pm 1.5 \pm 0.9) \times 10^{-5}$, and the inclusive branching fraction for $b \to s\gamma$ to be $(2.3 \pm 0.6 \pm 0.4) \times 10^{-4}$. These values are in agreement with the Standard Model prediction, and set bounds on the parameters of some extensions of the Standard Model. Other penguin decays such as $B \to \rho\gamma$ and $b \to s\ell^+\ell^-$ have been searched for but have not yet been seen.

There is also a CLEO measurement of $(1.8 \pm 0.6 \pm 0.2) \times 10^{-5}$ for the branching fraction of the sum of the decays $B^0 \to K^+\pi^-$ and $B^0 \to \pi^+\pi^-$. While the $K^+\pi^-$ final state is thought to occur primarily through the gluonic penguin diagram [Figure 1(d)], the $\pi^+\pi^-$ final state occurs primarily via a $b \to u$ tree level transition [Figure 1(a)]. The data favor equal branching ratios for the two decay modes, but only exclude either one being zero at a level of significance equivalent to about $2.5\sigma$.

Decays of $B$ mesons to two pseudoscalar mesons are particularly important for the study of CP violation. The $\pi^+\pi^-$ final state can be used to measure the angle $\alpha$ in the CKM triangle. Corrections for a penguin contribution to $B^0 \to \pi^+\pi^-$ can be made by making additional measurements of $B^+ \to \pi^+\pi^0$ and $B^0 \to \pi^0\pi^0$ and using isospin, or by measuring the rate asymmetry between $B^0 \to K^+\pi^-$ and $\bar{B}^0 \to K^-\pi^+$ and using SU(3) symmetry. It is likely that the CKM angle $\gamma$ will be measured using charged $B$ decays to $K^\pm\pi^0$, $\pi^\pm\pi^0$, $K^0\pi^\pm$ and $K^\pm\eta^{(\prime)}$, or using charged $B$ decays to $D^0 K^\pm$ final states.

We are looking forward to the measurement of additional rare $b$ decays such as $B \to \rho\gamma$ and $b \to s\ell^+\ell^-$. The Standard Model is already constrained by the existing measurements of $B^0$ mixing and $b \to s\gamma$. If deviations from the Standard Model are found in other radiative penguin decays, in CP asymmetry measurements, or in $K$ or $D$ meson decays, then possible extensions of the Standard Model must both explain the observed deviations and be consistent with the other measurements.

There are planned improvements to the CESR/CLEO symmetric $B$-factory, and asymmetric $B$ factories are under construction at KEK and SLAC [48], all with projected luminosities about ten times higher than that currently achieved at CESR/CLEO. In addition hadron collider $B$ experiments are being pursued (HERA-B, Tevatron, LHC)[74]. We hope that these efforts will lead to the observation of CP violation in the $B$ system, and that some evidence for non-Standard Model effects will be found.

# IX. Acknowledgements

We thank the National Science Foundation for support. We acknowledge informative conversations with A. Ali, G. Burdman, N. Deshpande, I. Dunietz, X. He, J. Hewett, N. Isgur, J. Rosner, A. Soni, L. Wolfenstein and our colleagues in the CLEO collaboration.

We apologize to all of those who do not appear in the above list, or who are omitted in the references below, but who have contributed significantly to the study of rare $b$ decays. It has been a constant struggle to finish this paper given the large number of theoretical preprints that appear every week, but it is a great joy to see so much activity in this field.